\documentclass[epj]{webofc}
\usepackage[varg]{txfonts}   % Web of Conferences font
\usepackage{epsfig}
\usepackage{graphics}
\wocname{EPJ Web of Conferences}
\woctitle{INPC 2013}
\begin{document}
\title{Rotation and alignment of high-$j$ orbitals in transfermium nuclei}
%
% subtitle is optional

\author{Xiao-Tao He \inst{1,2}\fnsep\thanks{\email{hext@nuaa.edu.cn}} \and
        Zhen-Hua Zhang\inst{3} \and
        Jin-Yan Zeng\inst{3}\and
        En-Guang Zhao\inst{4}\and
        Zhong-Zhou Ren\inst{5}\and
        Werner Scheid\inst{1}\and
        Shan-Gui Zhou\inst{4}
        % etc.
}

\institute{Institut f\"{u}r Theoretische Physik, Justus-Liebig-Universit\"{a}t
Giessen, Gie{\ss}en 35392, Germany
\and
           College of Material Science and Technology, Nanjing University of Aeronautics and Astronautics,
Nanjing 210016, China
\and
           State Key Laboratory of Nuclear Physics and Technology, School of Physics, Peking University, Beijing 100871, China
\and
           State Key Laboratory of Theoretical Physics, Institute of Theoretical Physics, Chinese Academy of Sciences, Beijing 100190, China
\and
           Department of Physics, Nanjing University, Nanjing 210093, China
          }

\abstract{%
 The structure of nuclei with $Z\sim100$ is investigated systematically by the Cranked Shell Model (CSM) with pairing correlations treated by a Particle-Number Conserving (PNC) method. In the PNC method, the particle number is conserved and the Pauli blocking effects are taken into account exactly. By fitting the experimental single-particle spectra in these nuclei, a new set of Nilsson parameters ($\kappa$ and $\mu$) is proposed. The experimental kinematic moments of inertia and the band-head energies are reproduced quite well by the PNC-CSM calculations. The band crossing, the effects of high-$j$ intruder orbitals and deformation are discussed in detail.
}
\maketitle
\section{Introduction}
\label{intro}

The exploration of the island of stability with high mass and charge, i.e., the region of superheavy elements (SHE), has been one of the fundamental questions in natural science. Great experimental progress has been made in synthesizing the superheavy elements. Up to now, superheavy elements with $Z\leq118$ have been synthesized via cold and hot fusion reactions~\cite{Hofmann00,Morita04,Oganessian07}. However, due to the extremely low production cross-sections, these experiments can rarely reveal the detailed spectroscopic information. One indirect way is to study lighter nuclei in the deformed region with $Z\approx100$ and $N\approx152$, which are the heaviest systems accessible in present in-beam experiments (see Refs.~\cite{Leino04,Herzberg04,herzberg08} and references therein). The strongly downsloping orbitals originating from the spherical subshells active in the vicinity of the predicted shell closures come close to the Fermi surface of transfermium nuclei due to deformation effect. The rotational properties of transfermium nuclei will be strongly affected by these spherical orbitals. The proton $1/2^{-}[521]$ orbital is of particular interest since it stems from the spherical $2f_{5/2}$ orbital. The spin$-$orbit interaction strength of $2f_{5/2}-2f_{7/2}$ partner governs the size of the possible $Z = 114$ shell gap. The Cranked Shell Model (CSM) with the pairing correlations treated by a Particle-Number Conserving (PNC) method~\cite{Zeng1983_NPA405-1,Zeng1994_PRC50-1388} is used to study the rotational and single-particle properties of $Z\sim100$ nuclei.

\section{Theoretical framework}
\label{sec:Formulism}

The Cranked Shell Model Hamiltonian of an axially symmetric
nucleus in the rotating frame is expressed as:
\begin{equation}
\ H_{\text{CSM}}=H_{0}+H_{\text{P}}=\sum_{n}(h_{\text{Nil}}-\omega j_{x})_{n}+H_{\text{P}}(0)+H_{\text{P}}(2)\ ,
\end{equation}
where $h_{\textrm{Nil}}$ is the Nilsson Hamiltonian\cite{Nilsson69},
$-\omega j_{x}$ is the Coriolis force with the cranking frequency $
\omega$ about the $x$ axis (perpendicular to the nuclear symmetry $z$ axis). $H_{\text{P}}$ is the pairing including monopole and quadrupole pairing correlations,
\begin{center}
\begin{eqnarray}
 \ H_{\text{P}}(0)
 &=&-G_{0}\sum_{\xi \eta }
 a_{\xi }^{\dagger}a_{\overline{\xi }}^{\dagger }a_{\overline{\eta }}a_{\eta }\ ,
 \\
 H_{\text{P}}(2)
 &=&-G_{2}\sum_{\xi \eta } q_{2}(\xi) q_{2}(\eta)
 a_{\xi}^{\dagger } a_{\overline{\xi}}^{\dagger}
 a_{\overline{\eta}}a_{\eta}\ ,
 \label{eq:Hp}
\end{eqnarray}
\end{center}
with $\overline{\xi }$ and $\overline{\eta }$ being the time-reversal
states of a Nilsson state $\xi$ and $\eta$, respectively. The quantity $q_{2}(\xi) =\sqrt{16\pi /5}\left\langle \xi \right\vert r^{2}Y_{20}\left\vert \xi \right\rangle $ is the
diagonal element of the stretched quadrupole operator, and $G_{0}$ and $G_{2}$ are the effective
strengths of monopole and quadrupole pairing interactions, respectively.

In our calculation, $h_{0}(\omega)=h_{\textrm{Nil}}-\omega j_{x}$ is diagonalized firstly to obtain the
cranked Nilsson orbitals. Then, $H_{\text{CSM}}$ is diagonalized in a
sufficiently large Cranked Many-Particle Configuration (CMPC) space
to obtain the yrast and low-lying eigenstates. Instead of
the usual single-particle level truncation in
common shell-model calculations, a cranked many-particle
configuration truncation (Fock space truncation)
is adopted which is crucial to make the PNC calculations
for low-lying excited states both workable and sufficiently
accurate~\cite{Wu89,Zhang13} . The eigenstate of $H_{\text{CSM}}$ is expressed as:
\begin{equation}
\left| \psi \right\rangle =\sum_{i}C_{i}\left| i\right\rangle \ ,
\label{eq:wf}
\end{equation}
where $\left\vert i\right\rangle $ is a cranked many-particle configuration
(an occupation of particles in the cranked Nilsson orbitals) and $C_{i}$ is the
corresponding probability amplitude.

The angular momentum alignment $\left\langle J_{x} \right\rangle$ of
the state $\left\vert \psi \right\rangle$ is given by:
\begin{equation}
 \left\langle \psi \right| J_{x}
 \left| \psi \right\rangle
 = \sum_{i}\left|C_{i}\right| ^{2}
   \left\langle i\right| J_{x}\left| i\right\rangle
 + 2\sum_{i<j}C_{i}^{\ast }C_{j}
   \left\langle i\right| J_{x}\left| j\right\rangle\ .
 \label{eq:Jx}
\end{equation}
The kinematic moment of
inertia is $J^{(1)}=\left\langle \psi \right\vert J_{x}\left\vert
\psi \right\rangle /\omega $.

\section{Results and discussions}
\label{sec:discussions}

The Nilsson parameters $(\kappa,\mu)$ proposed in
Refs.~\cite{Nilsson69,bengtsson85} cannot well describe the experimental
level schemes of transfermium nuclei while it is optimized to reproduce the experimental level
schemes for the rare-earth and actinide nuclei. By fitting the
experimental single-particle levels in the odd-$A$ nuclei with
$Z\approx100$, we obtained a new set of Nilsson parameters $\kappa$ and $\mu$ (see Table~\ref{tab-1})
which are dependent on the main oscillator quantum number
$N$ as well as on the orbital angular momentum $l$~\cite{zhang11,zhang12}.

\begin{table}[ht]
\centering
\caption{Nilsson parameters $\kappa$ and $\mu$ proposed for the nuclei
with $Z\approx100$. Taken from Ref.~\cite{zhang11,zhang12}.}
\label{tab-1}       % Give a unique label
% For LaTeX tables you can use
\begin{tabular}{llllllll}
\hline
$N$ & $l$ & $\kappa_{p}$ & $\mu_{p}$ & $N$ & $l$ & $\kappa_{n}$ & $\mu_{n}$\\\hline
4 & 0,2,4 & 0.0670 & 0.654 & & & & \\
5 & 1 & 0.0250 & 0.710 & 6 & 0 & 0.1600 & 0.320 \\
  & 3 & 0.0570 & 0.800 &   & 2 & 0.0640 & 0.200 \\
  & 5 & 0.0570 & 0.710 &   & 4,6 & 0.0680 & 0.260 \\
6 & 0,2,4,6 & 0.0570 & 0.654 & 7 & 1,3,5,7 & 0.0634 & 0.318 \\\hline
\end{tabular}
% Or use
%\vspace*{5cm}  % with the correct table height
\end{table}

\begin{figure}[hb]
\centering
\sidecaption
\includegraphics[width=7cm]{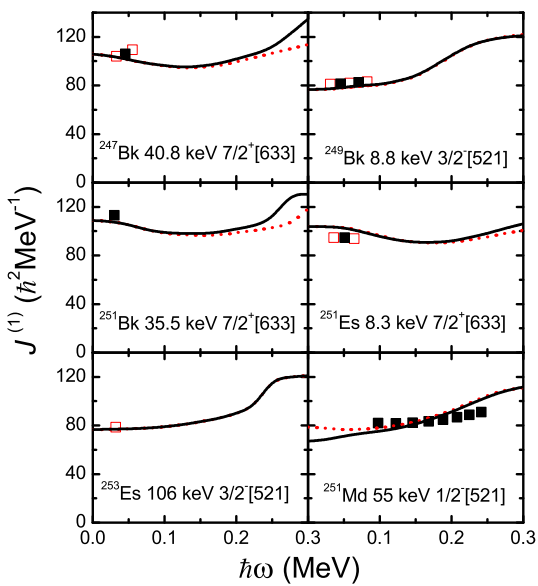}
\caption{The experimental and
calculated MOI's $J^{(1)}$ of the excited 1-qp bands in odd-$Z$ Bk, Es, and Md isotopes.
The data are taken from Refs.~\cite{Herzberg04,herzberg08} and references therein.
The experimental MOI's are denoted by full squares
(signature $\alpha=+1/2$) and open squares (signature $\alpha=-1/2$), respectively.
The calculated MOI's by the PNC method are denoted by solid lines
(signature $\alpha=+1/2$) and dotted lines (signature $\alpha=-1/2$), respectively.
The effective pairing interaction
strengths for both protons and neutrons for all these odd-$N$ nuclei are,
$G_n=0.30$~MeV, $G_{2n}=0.02$~MeV, $G_p=0.25$~MeV, and $G_{2p}=0.01$~MeV. Taken from Ref.~\cite{zhang11}.}
\label{fig:MOIoddZE}       % Give a unique label
\end{figure}

The CMPC space in
the work of Ref.~\cite{zhang12} is constructed
in the proton $N=4, 5, 6$ shells and the neutron $N=6, 7$ shells.
The dimensions of the CMPC space for the nuclei with $Z \approx 100$
are about 1000 both for protons and neutrons. Figure~\ref{fig:MOIoddZE} gives
the experimental and calculated moments of inertia of excited 1-qp bands in the
odd-$Z$ Bk, Es, and Md isotopes (taken from Ref.~\cite{zhang12}). The
data are well reproduced by the PNC calculations.
Only one signature band was observed in $^{251}$Md. We calculated $J^{(1)}$'s
for two signature partner bands which vary
smoothly with frequency in $^{251}$Md.

\begin{figure}[htb]
\centering
\sidecaption
\includegraphics[width=7cm,clip]{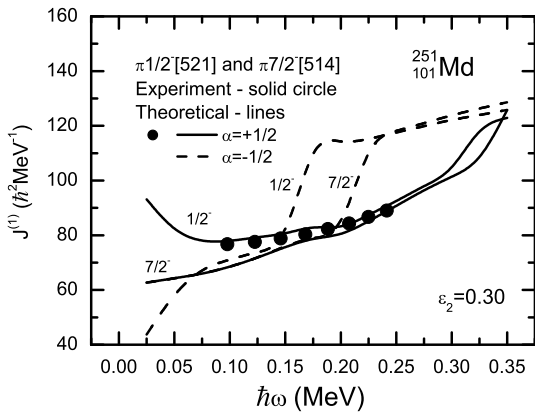}
\caption{Experimental and calculated kinematic moment of inertia
$J^{(1)}$ of the $\pi1/2^{-}[521]$ and $\pi7/2^{-}[514]$ bands in
$^{251}$Md. The experimentally observed $1/2^{-}[521](\alpha=+1/2)$
band is denoted by solid circles. Solid and dashed lines are used
for the calculated $\alpha=+1/2$ and $\alpha=-1/2$ bands,
respectively. Taken from Ref.~\cite{He09}.}
\label{fig:J1-Md}       % Give a unique label
\end{figure}

In order to investigate the effect of the proton $N=7$ shell on the rotational
properties of the transfermium nuclei, the proton $N=7$ shell is included to construct
the CMPC space~\cite{He09}. We find that the $1/2^{-}[770]$ orbital plays an important
role in the rotational properties of $^{251}$Md. Figure~\ref{fig:J1-Md} (taken from Ref.~\cite{He09}) shows the experimental and calculated kinematic moment of inertia $J^{(1)}$ of the $1/2^{-}[521]$ band in $^{251}$Md. A sharp backbending of the $\alpha=-1/2$ band takes place at a very low frequency ($\hbar\omega\approx0.15 $MeV) while the $\alpha=+1/2$ band varies smoothly in the whole observed frequency range. The signature splitting is due to the band crossing between the $1/2^{-}[521]$ and $1/2^{-}[770]$ configurations at $\hbar\omega\approx0.15$ MeV for $\alpha=-1/2$ and $\hbar\omega\approx0.30$ MeV for $\alpha=+1/2$.
Since the position of the $1/2^{-}[770]$ orbital is very sensitive to the deformation~\cite{chasman77}, we calculate $^{251}$Md for $\varepsilon_{2}=0.28$ and $0.255$ with and without the proton $N=7$ shell, respectively. There is no signature splitting when the proton $N=7$ shell is not included whether we take $\varepsilon_{2}=0.28$ or $\varepsilon_{2}=0.255$. The signature splitting occur at $\hbar\omega\approx0.225$ ($\hbar\omega\approx0.275$) for $\varepsilon_{2}=0.28$ ($\varepsilon_{2}=0.255$) when the effect of the proton $N=7$ shell is considered~\cite{He13}.

\section{Summary}
\label{sec:summary}

The rotational bands in the nuclei with $Z\approx100$ are investigated by using a Cranked Shell Model (CSM) with the pairing correlations treated by a Particle-Number Conserving (PNC) method. In the PNC-CSM method, the blocking effects are taken into account exactly. By fitting the experimental single-particle spectra in these nuclei, a new set of Nilsson parameters ($\kappa$ and $\mu$) is proposed. The experimentally observed variations of moment of inertia for these nuclei with the frequency $\omega$ are reproduced very well by the PNC-CSM calculations. The high-$j$ intruder proton orbital $\pi 1j_{15/2}$ ($1/2^{-}[770]$) plays an important role in the sharp backbending of the $1/2^{-}[521] \alpha=-1/2$ band for $^{251}$Md.

%
% BibTeX or Biber users please use (the style is already called in the class, ensure that the "woc.bst" style is in your local directory)
% \bibliography{name or your bibliography database}

\end{document}